\documentclass[a4paper,11pt]{article}
\usepackage{pos}
\usepackage{hyperref}
\usepackage{wasysym}
\usepackage{caption}
\usepackage{subcaption}
\usepackage{siunitx}
\usepackage{color}
\usepackage{mathtools}
\usepackage{lineno}

\DeclarePairedDelimiterXPP\BigOSI[2]%
  {\mathcal{O}}{(}{)}{}%
  {\SI{#1}{#2}}
\newcommand{\dd}{\mathrm{d}}

\title{Search for High-Energy Neutrinos from TDE-like Flares with IceCube}
 \ShortTitle{Neutrinos and TDEs}

\author{The IceCube Collaboration \\{\normalsize \normalfont(a complete list of authors can be found at the end of the proceedings)}\\}

\emailAdd{jannis.necker@desy.de}

\abstract{The collected data of IceCube, a cubic kilometer neutrino detector array in the Antarctic ice, reveal a diffuse flux of astrophysical neutrinos. The extragalactic sources of the majority of these neutrinos however have yet to be discovered. Tidal Disruption Events (TDEs), disruption outbursts from black holes that accrete at an enhanced rate, are candidates for being the sources of extragalactic, high-energy neutrinos. Stein at al. (2021)\cite{steinTidalDisruptionEvent2021b} and Reusch et al. (2022) \cite{reuschCandidateTidalDisruption2022a} have reported the coincidence of two likely TDEs from supermassive black holes and public IceCube neutrino events (alerts). Further work by van Velzen et al. (2021) identified a third event in coincidence with a high-energy neutrino alert and a $3.7 \sigma$ correlation between a broader set of similar TDE-like flares and IceCube alerts. We conducted a stacking analysis with a 29-flare subset of the TDE-like flares tested by van Velzen et al. This work was done with neutrinos with energies above $\BigOSI{100}{\GeV}$. The resulting p-value of 0.45 is consistent with background. In this contribution I will discuss the results of the stacking analysis as well as the impact of using different reconstruction algorithms on the three correlated realtime alerts.

\vspace{4mm}
{\bfseries Corresponding authors:}
Jannis Necker$^{1*}$\\
{$^{1}$ \itshape DESY}\\[4mm]
$^*$ Presenter

\ConferenceLogo{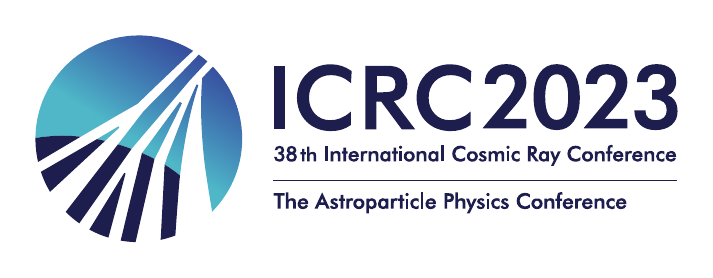}

\FullConference{The 38th International Cosmic Ray Conference (ICRC2023)\\ 26 July -- 3 August, 2023\\ Nagoya, Japan}

}

\begin{document}
\maketitle

\section{Introduction}
IceCube is a cubic kilometer scale neutrino detector located at the South Pole. It consists of 5160 Digital Optical Modules (DOMs) that detect high-energy astrophysical neutrinos by measuring Cerenkov light of secondary particles produced by the interaction of a neutrino. The charged-current interaction of a muon neutrino will produce a muon that can travel for kilometers in the ice. This will leave an elongated light pattern in the detector, dubbed muon track (see, for example, Figure \ref{fig:lancel_view}), that allows for a reconstruction of the arrival direction with an uncertainty of $\BigOSI{1}{\degree}$. For selected, promising high-energy neutrino events, IceCube sends out an alert in realtime through the \textit{Global Coordinates Network}\footnote{\url{https://gcn.nasa.gov}} (GCN).

One of the observatories following up these notifications with observations is the \textit{Zwicky Transient Facility} (ZTF). It is an optical telescope with a large field-of view on Mount Palomar in California. For neutrino events that are sufficiently well localized and have a probability high enough for being of astrophysical origin, ZTF systematically observes the neutrino's localization region \cite{steinNeutrinoFollowupZwicky2023}.

One of the candidate high-energy neutrino source populations that this follow-up program is sensitive to are \textit{Tidal Disruption Events} (TDEs), rare astrophysical transient events that happen, when a star passes close to a \textit{Supermassive Black Hole} (SMBH) that are believed to reside in the center of almost every galaxy. The star can disintegrate, and the resulting stellar debris forms an accretion disk that emits radiation across the electromagnetic spectrum. TDEs have been suggested as the sources of high-energy neutrinos and, indeed, ZTF has identified three likely TDEs in the footprint of high-energy neutrino events.

The reports of these coincidences are reviewed in Section \ref{sec:intro_coincidences}. Section \ref{sec:stacking} describes a stacking analysis we performed to investigate the correlation between likely TDEs and high-energy neutrinos, before taking a closer look at the three alert events in Section \ref{sec:resim} and drawing conclusions in Section \ref{sec:conclusions}.

\section{Coincidences between High-Energy Neutrino Alerts and Accretion Flares}
\label{sec:intro_coincidences}

\begin{figure}
    \centering
    \begin{subfigure}[b]{0.29\textwidth}
        \centering
        \includegraphics[width=\textwidth]{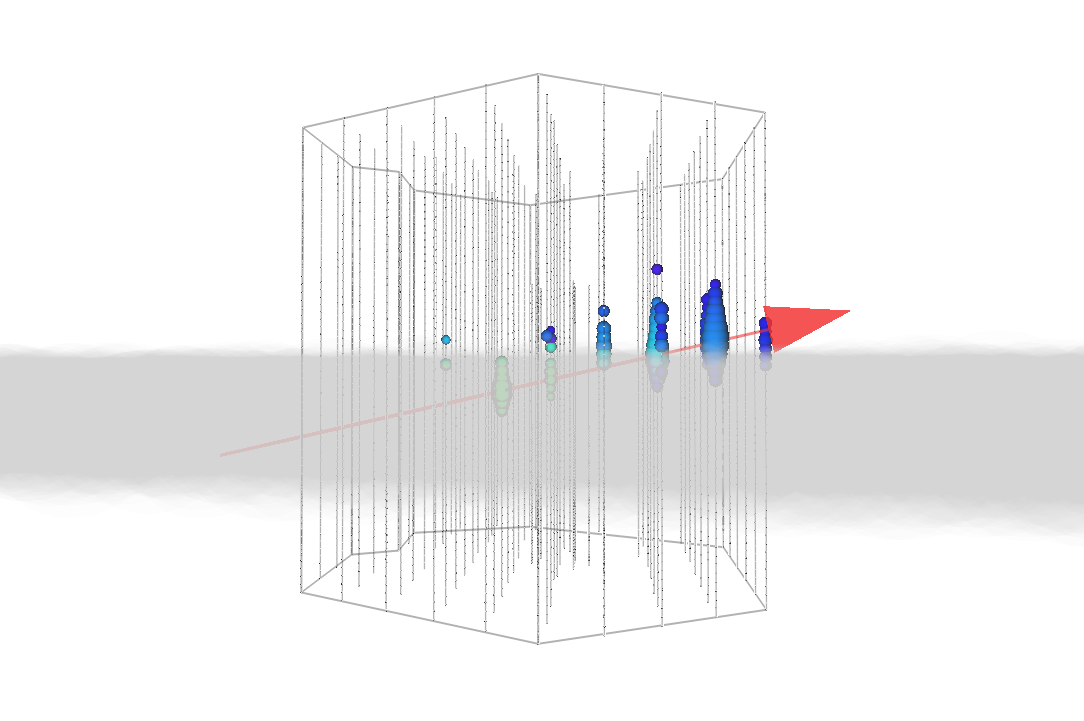}
        \caption{IceCube-191009A}
        \label{fig:bran_view}
    \end{subfigure}
    \begin{subfigure}[b]{0.29\textwidth}
        \centering
        \includegraphics[width=\textwidth]{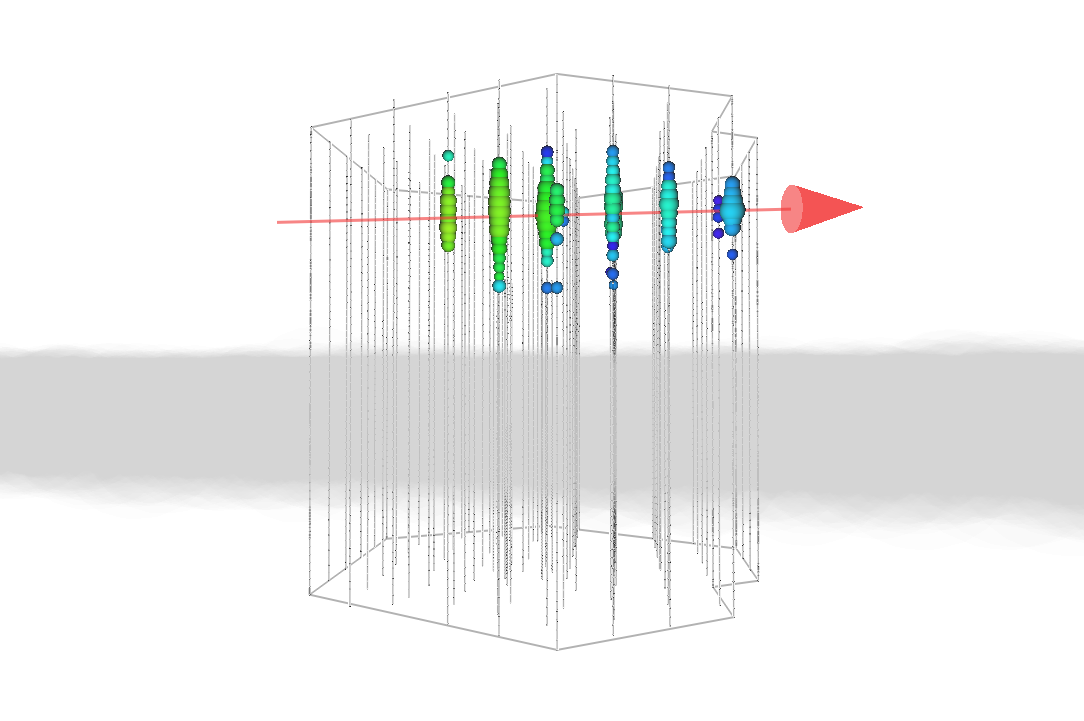}
        \caption{IceCube-191119A}
        \label{fig:lancel_view}
    \end{subfigure}
    \begin{subfigure}[b]{0.29\textwidth}
        \centering
        \includegraphics[width=\textwidth]{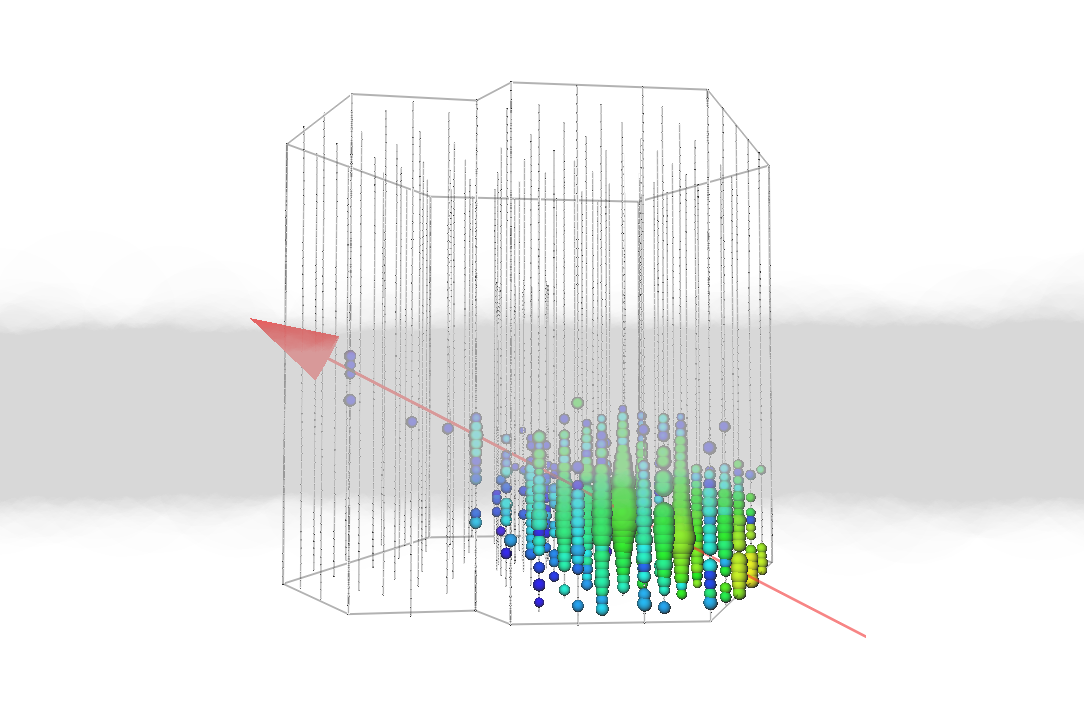}
        \caption{IceCube-200530A}
        \label{fig:tywin_view}
    \end{subfigure}

    \caption{Event views of the three neutrino alert events that were reported to be coincident with TDEs or Accretion Flares.}
\end{figure}

In 2019 on October 1st, IceCube reported the detection of the high-energy neutrino IceCube-191001A (Figure \ref{fig:bran_view}). The TDE AT2019dsg was reported by ZTF to be a candidate counterpart \cite{steinTidalDisruptionEvent2021b}. Again, In 2020 on May 30th, ZTF reported the candidate TDE AT2019fdr to be coincident with the localization region of the high-energy neutrino IceCube-200530A (Figure \ref{fig:tywin_view}) \cite{reuschCandidateTidalDisruption2022a}. Both of these optically detected flares show a delayed flare in the infrared. This can be interpreted as a dust echo, emission by hot dust, that gets heated by the forming accretion disk. A systematic study compiled a sample of flares similar to AT2019fdr and AT2019dsg, dubbing them accretion flares because most of them are not unambiguously classified TDEs, but do present a state of enhanced accretion onto a SMBH. Correlating this sample with archival high-energy neutrino alerts, the study found a third coincidence, AT2019aalc and IceCube-191119A (see Figure \ref{fig:lancel_view}). Furthermore, the sample of accretion flares is claimed to be correlated with the sample of high-energy neutrino alerts at a level of $3.7 \sigma$. \cite{vanvelzenEstablishingAccretionFlares2021}

\section{Stacking Analysis}
\label{sec:stacking}

We performed an unbinned maximum likelihood analysis to look for correlation between accretion flares compiled by \cite{vanvelzenEstablishingAccretionFlares2021} and a more extensive neutrino dataset using the method outlined in \cite{theicecubecollaborationMultimessengerObservationsFlaring2018}.

We use a selection of nine years of track-like muon events that is optimized for point-source searches
\cite{aartsenAllskySearchTimeintegrated2017} and extends from May 2011 up until May 2020. The estimated proxy for the muon energy goes from \SI{100}{\GeV} to \SI{10}{\peta\eV}.

From the catalog of accretion flares \cite{vanvelzenEstablishingAccretionFlares2021}, we select the sources with a strong dust echo. The strength of this echo is defined as the increase in infrared flux divided by the pre-flare variability:
\begin{equation}
    S = \frac{\Delta F_\mathrm{IR}}{F_\mathrm{rms}}.
\end{equation}
For a dust echo strength of $S > 10$ the flare is more likely to be due to a TDE than to regular AGN variability \cite{vanvelzenEstablishingAccretionFlares2021}. From the original sample of 63 flares used in the correlation analysis, we use a subselection of the 29 sources with $S > 10$. This does include the three flares that were reported to be coincident with high-energy neutrino alerts.

To build the likelihood, we follow the standard point source likelihood approach. Here, we use a variant where we leave the weights of the individual sources as free parameters in the fit \cite{abbasiConstrainingHighenergyNeutrino2023}. The likelihood can be written as

\begin{equation}
\label{eq:stack_llh_fit}
    \mathcal{L}(\vec{n_{\mathrm{s}}}, \gamma) = \prod _{\mathrm{i=0}} ^\mathrm{N} \left[ \frac{1}{\mathrm{N}} \sum_\mathrm{j=0} ^\mathrm{M} n_\mathrm{s,j} \cdot \mathcal{S}_\mathrm{j}(\theta_\mathrm{i}, \gamma)  + \left( 1- \frac{n_{\mathrm{s}}}{\mathrm{N}} \right) \mathcal{B}(\theta_\mathrm{i})  \right]
\end{equation}
where, for each source $j$, the signal PDF $\mathcal{S}_j$ is evaluated for each neutrino $i$ with the properties $\theta_i$ and weighted with the relative source weight $n_{s, j}$, indicating the number of signal neutrinos from source j. The background PDF is given by $\mathcal{B}$. $n_s$ denotes the total number of signal neutrinos $n_s = \sum _{\mathrm{j}=0} ^\mathrm{M} n_{s, j}$.

Both the signal and the background PDF can be described as a product of a spatial, temporal and energy part.
The dataset is dominated by background, which mostly comes from atmospheric neutrinos. We can thus assume the background time PDF to be uniform and model the declination dependent energy and spatial part from the data.
The spatial component of the signal PDF can be described as a two-dimensional Gaussian distribution centered at the reconstructed arrival direction and with the width equal to the angular uncertainty $\sigma$ of each event.
It is important to point out that the angular uncertainties used here differ from the uncertainty contours reported for the alerts and are typically much smaller. Because the best-fit position of IceCube-191001A and IceCube-191119A is $\BigOSI{1}{\degree}$ away from the position of the respective accretion flare, we do not expect them to contribute significantly in our stacking analysis although they are present in the dataset (IceCube-200530A happened one day after our dataset end). The difference between the uncertainties is under investgation (see Section \ref{sec:resim}).
The energy component depends on the assumed spectral index of the sources and on the declination of each individual source, and is obtained from Monte-Carlo simulations. For the temporal component, we assume a 1-year box model, starting at the peak of the optical lightcurve. This is motivated by the three coincidences of high-energy neutrino alerts and accretion flares.

\begin{figure}
	\centering
	\includegraphics[width=0.5\textwidth]{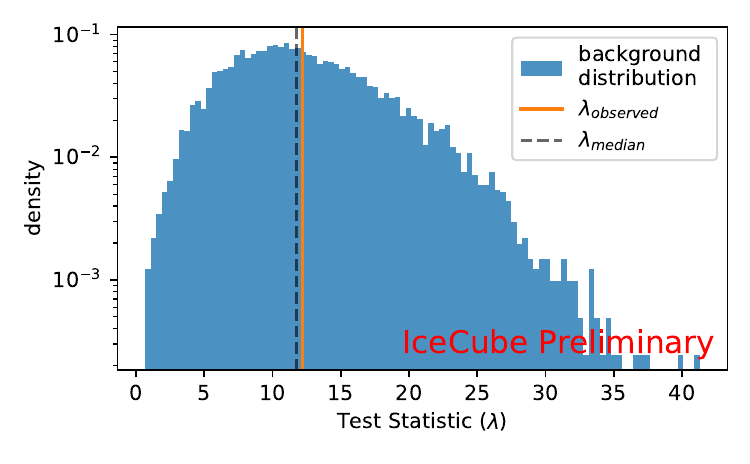}
	\caption{Distribution of background test statistic values and the observed value. }
	\label{fig:stacking_res}
\end{figure}

We compare the signal hypothesis to the background hypothesis by defining our test statistic as
\begin{equation}
	\lambda = 2 \cdot \frac{\mathcal{L}(\vec{n_s}, \gamma)}{\mathcal{L}(n_s=0)}.
\end{equation}
We run background trials by calculating $\lambda$ for a pseudo-data set, created by scrambling the data in Right Ascension and shuffling the event times. To estimate the significance of the best fit test statistic value, we compare it to the test statistic distribution obtained by running $10^4$ background trials. This background distribution is shown in Figure \ref{fig:stacking_res}. The resulting p-value is $p=0.45$ which is fully compatible with background.

\begin{figure}
    \centering
    \includegraphics[width=0.5\textwidth]{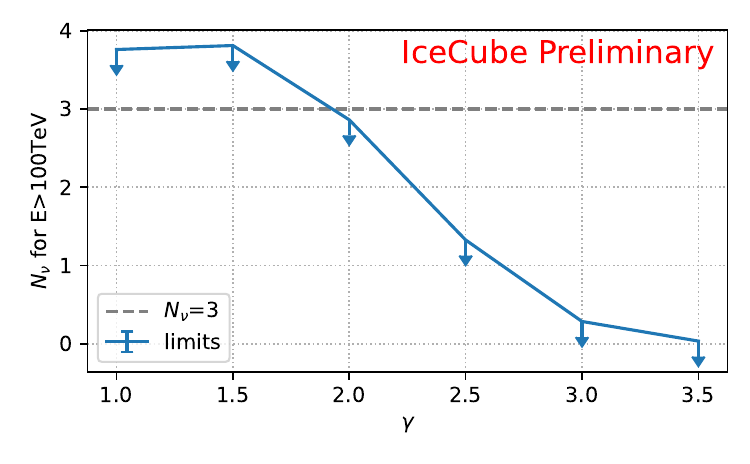}
    \caption{Estimation of the number of alert events allowed by the upper limit from the stacking analysis.}
    \label{fig:stacking_alert_estimation}
\end{figure}

This non-detection can still be compatible with the claimed correlation with high-energy neutrino alerts if the sources have a spectral index that is hard enough that the flux at high energies produces the three observed alert neutrinos but below the sensitivity of the stacking analysis in this work at lower energies. To quantify this, we estimate the number of alert events compatible with our stacking analysis. We calculate the expected number of neutrino events in our data sample with a reconstructed energy of $E_\nu > \qty{100}\TeV$, given the 90\% upper limit on the flux from the 29 sources in our analysis. This is shown as a function of the spectral index in Figure \ref{fig:stacking_alert_estimation}. A detection of three neutrino alert events associated with the accretion flare sample seems reasonable if the neutrino emission of these sources follows a hard spectrum with a spectral index of $\gamma < 2$.

\section{Investigating Alert coincidences}
\label{sec:resim}

Because the result of the stacking analysis is compatible with background, it is important to take a look at the alert events that contribute to the result of the correlation analysis,  concerning the correlation between accretion flares and IceCube high-energy neutrino alerts \cite{vanvelzenEstablishingAccretionFlares2021}. This study relies on the uncertainty regions published in realtime.

The first attempt to produce an uncertainty region for a neutrino event that was reported over the real time stream was done for IceCube-160427A after the report of a possibly associated young supernova \cite{kankareSearchTransientOptical2019a}. The event was reconstructed using the \textit{Millipede} \cite{aartsenEnergyReconstructionMethods2014} algorithm and the surrounding area was scanned on a grid, producing a likelihood map. Uncertainty areas were derived from this map using likelihood values obtained from simulations of the alert event.

Since then, the likelihood values derived from the simulations of IceCube-160427A were used to produce the uncertainty regions for all following high-energy neutrino alerts, with the exception of IceCube-170922A. Producing simulations also for other neutrino events, allows for an event-by-event estimation of the uncertainty region. IceCube investigations have shown in the past that the coverage of the regions derived from IceCube-160427A differ from the claimed, possibly due to systematic uncertainties \cite{lagunasgualdaStudiesSystematicUncertainty2021}.

This new approach relies on a robust framework that produces simulations of events that are similar to the alert events. Previously, similar events were defined by a difference in true neutrino direction of no more than 2 degrees, a distance of the corresponding muon track of no more than $\qty{50}\m$ and a total charge within 20\% of the measured event \cite{lagunasgualdaStudiesSystematicUncertainty2021} (see Table \ref{tab:schemes}). This has two main caveats:

First, many neutrinos have to be simulated to get similar muon events. In particular, the photon propagation will be performed for events that will not be selected as similar events based on deposited charge (see step 7 in Table \ref{tab:schemes}). Second, there is no handle on the actual shape of the event in the detector. Events that do not deposit light on their two far ends in the detector will have worse angular resolution. So it is important to also reproduce this pattern in the simulations.

\begin{table}
    \begin{tabular}{l | l | l}
        Step & Old & New \\
        \hline
        1: \textbf{Simulation} & $\nu$, its interaction and $\mu$ production & Place $\mu$ in detector  \\
        2: \textbf{Selection} & $\mu$ within $\qty{50}\m$ of original track & - \\
        3: \textbf{Simulation} & \multicolumn{2}{c}{ $\mu$ propagation \& energy losses} \\
        4: \textbf{Selection} & - &  $\mu$ with similar energy deposition patterns \\
        5: \textbf{Simulation}& \multicolumn{2}{c}{ ice systematics, photon propagation and detector response} \\
        6: \textbf{Selection} & $\mu$ within 20\% deposited charge & -
    \end{tabular}
    \caption{Different steps in the old and new scheme for simulating similar events of high-energy neutrino alert events.}
    \label{tab:schemes}
\end{table}

To address these, we can estimate the energy deposition of the alert event and compare this to the simulated energy deposition of the simulated muon. This allows for a selection of similar events \textit{before} the expensive step of photon propagation. If we can do this not only for the total energy but for segments along the track, this also gives us a handle on the shape of the event.

\textit{Truncated Energy} \cite{abbasiImprovedMethodMeasuring2013} is a well established algorithm within IceCube that aims at determining the energy of the incident muon by calculating the specific energy loss $\dd E/ \dd x$ along the muon's path given a track hypothesis. The track is divided into several fixed length segments and the specific energy loss is calculated individually for each of the segments. Multiplying by the segment length of $l=\qty{120}\m$, we get the deposited energy in the segment $i$ as
\begin{equation}
    \label{eq:e_meas}
    E_i = \left( \frac{\dd E}{\dd x} \right)_i \cdot l.
\end{equation}
This can be compared to the simulated energy deposition in the same segments.

\begin{figure}
    \centering
    \begin{subfigure}[b]{0.4\textwidth}
        \centering
        \includegraphics[width=\textwidth]{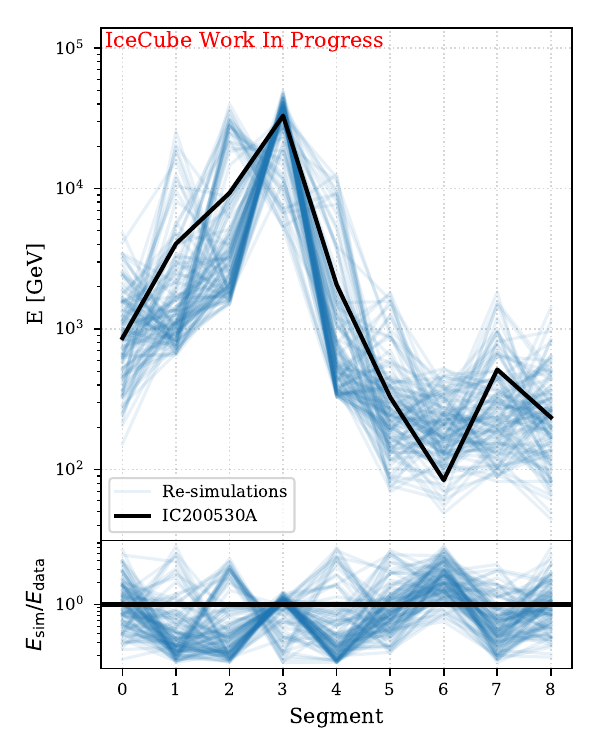}
        \caption{IceCube-200530A}
        \label{fig:tywin_sim}
    \end{subfigure}
    \begin{subfigure}[b]{0.4\textwidth}
        \centering
        \includegraphics[width=\textwidth]{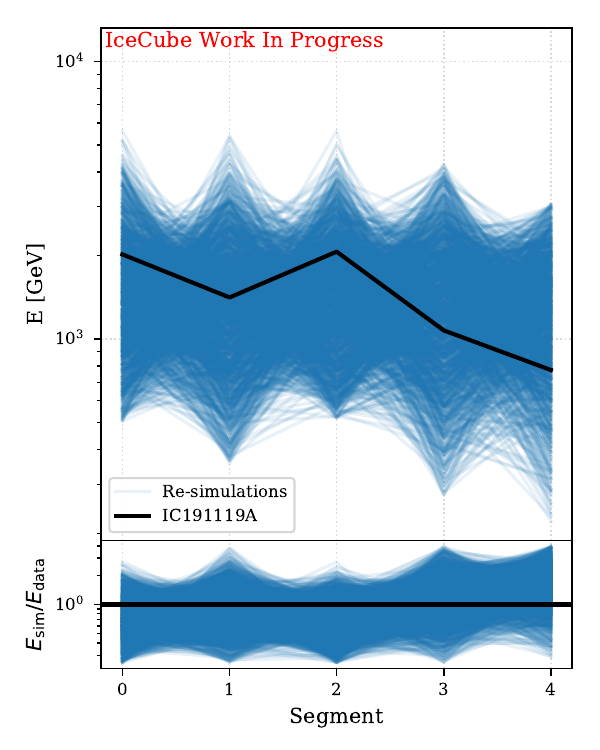}
        \caption{IceCube-191119A}
        \label{fig:lancel_sim}
    \end{subfigure}

    \caption{The simulated and measured energy depositions in the detector in segments.}
    \label{fig:sim}
\end{figure}

\begin{figure}
    \centering
    \begin{subfigure}[b]{0.49\textwidth}
        \centering
        \includegraphics[width=\textwidth]{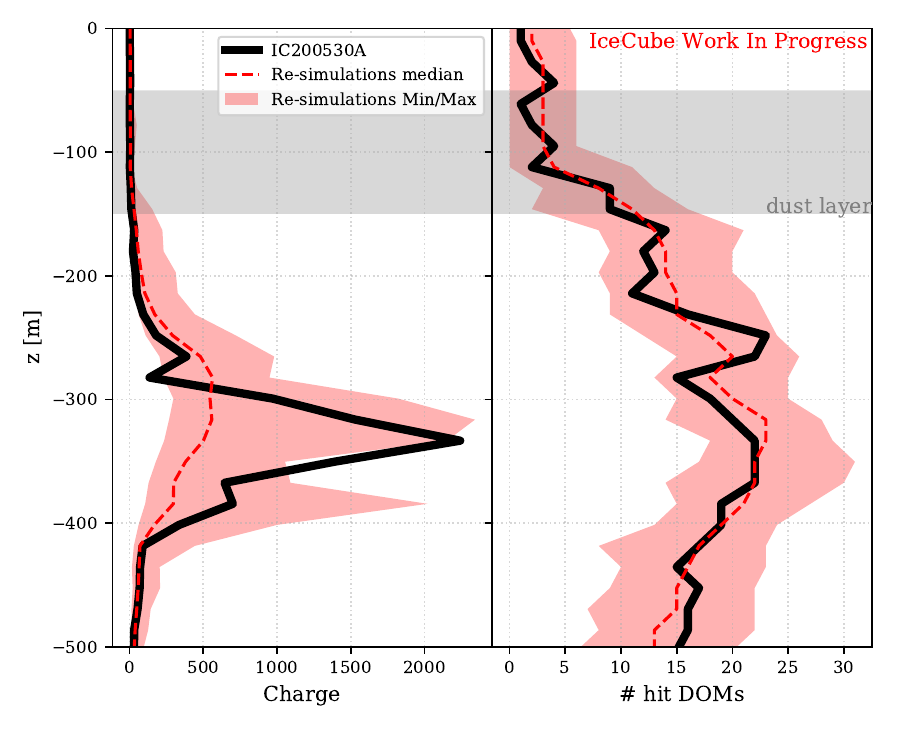}
        \caption{IceCube-200530A}
        \label{fig:tywin_charge}
    \end{subfigure}
    \begin{subfigure}[b]{0.49\textwidth}
        \centering
        \includegraphics[width=\textwidth]{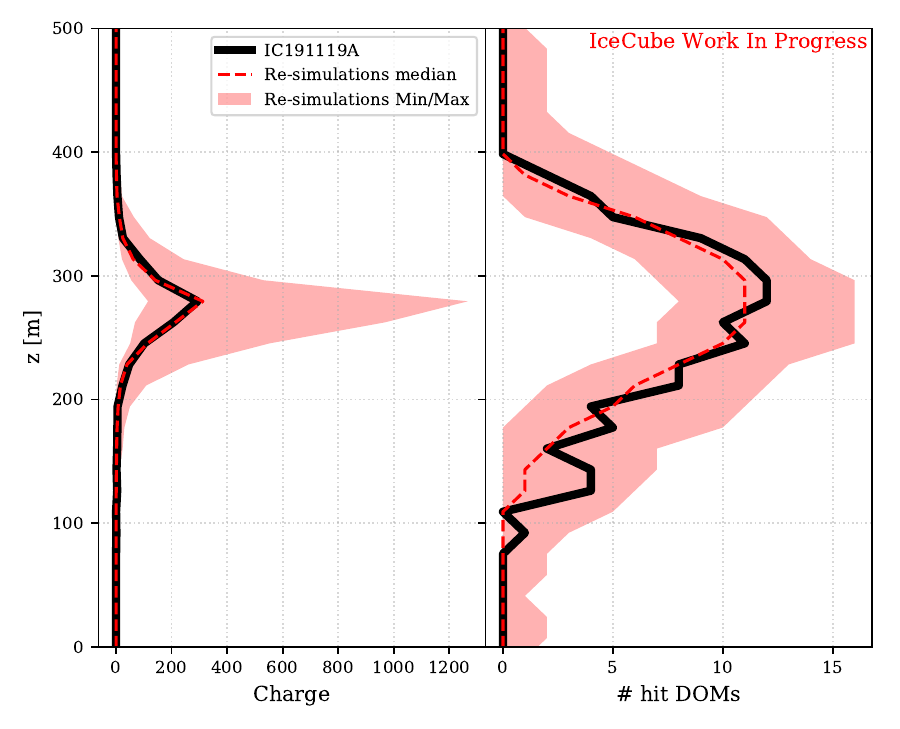}
        \caption{IceCube-191119A}
        \label{fig:lancel_charge}
    \end{subfigure}
    
    \begin{subfigure}[b]{0.49\textwidth}
        \centering
        \includegraphics[width=\textwidth]{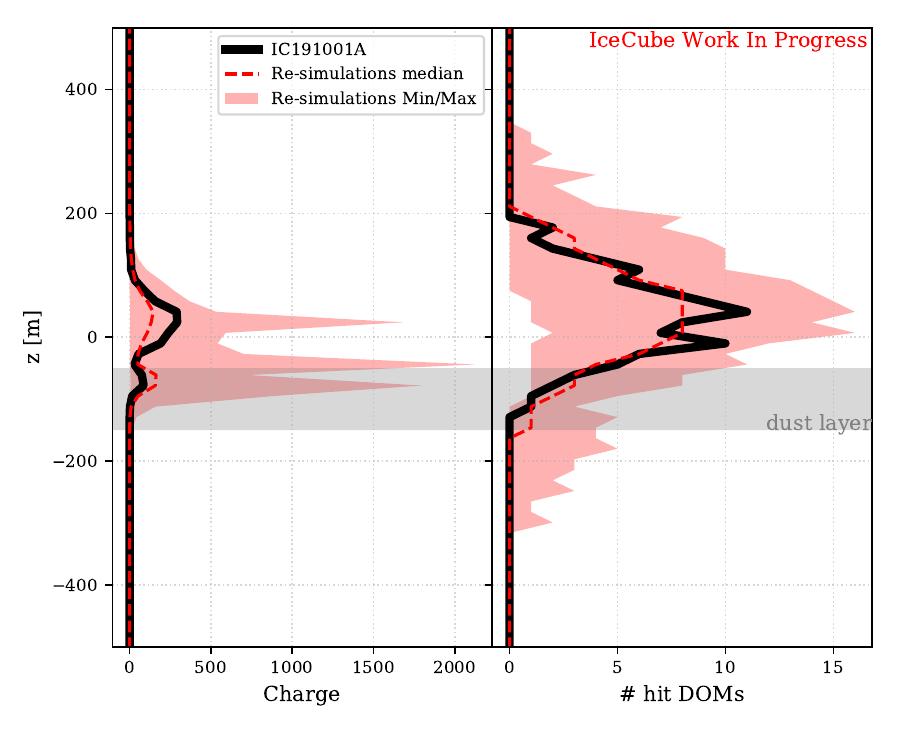}
        \caption{IceCube-191009A}
        \label{fig:bran_charge}
    \end{subfigure}
    \begin{subfigure}[b]{0.49\textwidth}
        \centering
        \includegraphics[width=\textwidth]{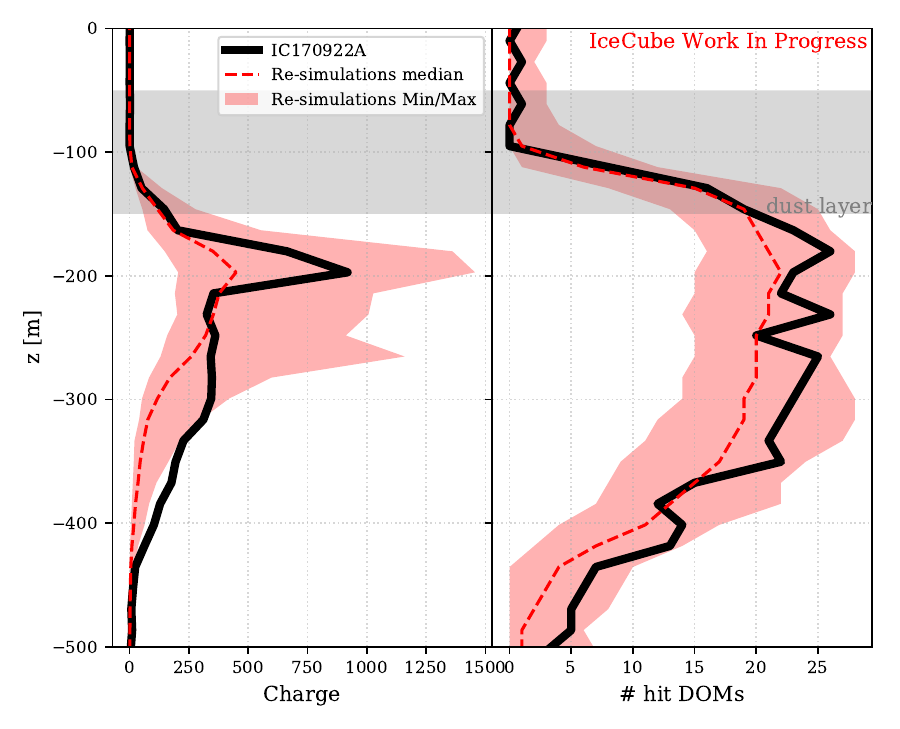}
        \caption{IceCube-170922A}
        \label{fig:txs_charge}
    \end{subfigure}

    \caption{Distribution of measured charge and number of hit DOMs as a function of height in the detector.}
    \label{fig:charge}
\end{figure}

The measured energy deposition is shown in Figure \ref{fig:sim} for the two through-going track events IceCube-200530A and IceCube-191119A, along with the simulated energy depositions produced with the new simulation scheme (see Table \ref{tab:schemes}). Apparently, the simulated energy losses can match the measured energy depositions. To verify that these similarities persist at detector level, Figure \ref{fig:charge} shows the distribution of charge and number of hit DOMs as a function of the height in the detector. For IceCube-200530A, IceCube-191119A and IceCube-170922A we show the results from the new simulation scheme.
IceCube-191119A (Figure \ref{fig:lancel_charge}) is perfectly re-produced by the mean of the simulations because of its smooth energy depositions along its path in the detector (see Figure \ref{fig:lancel_view}).
IceCube-200530A (Figure \ref{fig:tywin_charge}) has a peculiar energy deposition pattern, as mentioned already above. It deposits a lot of energy shortly after entering the detector and very little later on along the track (see Figures \ref{fig:tywin_view} and \ref{fig:tywin_sim}). This pattern is an outlier according to the simulations but is reproduced in some more extreme cases.
IceCube-170922A is a long, through-going track, and its charge depositions are also well reproduced by the simulations.
IceCube-191001A is a starting track (see Figure \ref{fig:bran_view}), where the interaction happens inside the detector volume and can not be simulated with the new scheme. Instead, we verify that the simulations obtained with the old scheme reproduce this event well (see Figure \ref{fig:bran_charge}).

In summary, this validates that there are now simulations for all three of the neutrino events reported to be coincident with an accretion flare. Using their reconstructed arrival directions will allow us to re-calibrate the likelihood contours of the alert events to test their robustness and eventually the robustness of the associations with the accretion flares. First results show that this distribution obtained from simulations is narrower than the distance to the respective optical flare, but more detailed studies are ongoing.

It is important to note that the presented analysis refers specifically to the three mentioned alert events, and the conclusions can not be transferred to other neutrino events.

\section{Conclusions}
\label{sec:conclusions}

We followed up on the report of a correlation between high-energy neutrino alerts and Accretion Flares. We conducted a stacking analysis to investigate this correlation with the full  IceCube data sample, allowing us to probe also much lower energies than the high-energy alert events. The result is compatible with background. To allow both the reported correlation with high-energy neutrino alerts and the non-detection using the full IceCube data sample, the sources have to have a spectral index of $\gamma < 2$.
To investigate the reported coincidences of high-energy neutrino alerts and accretion flares, we present an updated scheme of re-simulating the neutrino events and show that the energy deposition patterns are reproduced well. We will use them in the future to test the reported contours of the high-energy neutrino alerts for their robustness.

\bibliography{accretion_flare_stacking}
\bibliographystyle{ICRC}

%
%
%
\section*{Full Author List: IceCube Collaboration}

\scriptsize
\noindent
R. Abbasi$^{17}$,
M. Ackermann$^{63}$,
J. Adams$^{18}$,
S. K. Agarwalla$^{40,\: 64}$,
J. A. Aguilar$^{12}$,
M. Ahlers$^{22}$,
J.M. Alameddine$^{23}$,
N. M. Amin$^{44}$,
K. Andeen$^{42}$,
G. Anton$^{26}$
% ...
\\
\\
% ...
$^{12}$ Universit{\'e} Libre de Bruxelles, Science Faculty CP230, B-1050 Brussels, Belgium \\
% ...
$^{17}$ Department of Physics, Loyola University Chicago, Chicago, IL 60660, USA \\
$^{18}$ Dept. of Physics and Astronomy, University of Canterbury, Private Bag 4800, Christchurch, New Zealand \\
% ...
$^{22}$ Niels Bohr Institute, University of Copenhagen, DK-2100 Copenhagen, Denmark \\
$^{23}$ Dept. of Physics, TU Dortmund University, D-44221 Dortmund, Germany \\
% ...
$^{26}$ Erlangen Centre for Astroparticle Physics, Friedrich-Alexander-Universit{\"a}t Erlangen-N{\"u}rnberg, D-91058 Erlangen, Germany \\
% ...
$^{40}$ Dept. of Physics and Wisconsin IceCube Particle Astrophysics Center, University of Wisconsin{\textendash}Madison, Madison, WI 53706, USA \\
% ...
$^{42}$ Department of Physics, Marquette University, Milwaukee, WI, 53201, USA \\
% ...
$^{44}$ Bartol Research Institute and Dept. of Physics and Astronomy, University of Delaware, Newark, DE 19716, USA \\
% ...
$^{63}$ Deutsches Elektronen-Synchrotron DESY, Platanenallee 6, 15738 Zeuthen, Germany  \\
$^{64}$ Institute of Physics, Sachivalaya Marg, Sainik School Post, Bhubaneswar 751005, India \\
% ...

\subsection*{Acknowledgements}

\noindent
The authors gratefully acknowledge the support from the following agencies and institutions:
USA {\textendash} U.S. National Science Foundation-Office of Polar Programs,
U.S. National Science Foundation-Physics Division,
U.S. National Science Foundation-EPSCoR,
Wisconsin Alumni Research Foundation,
Center for High Throughput Computing (CHTC) at the University of Wisconsin{\textendash}Madison,
Open Science Grid (OSG),
Advanced Cyberinfrastructure Coordination Ecosystem: Services {\&} Support (ACCESS),
Frontera computing project at the Texas Advanced Computing Center,
U.S. Department of Energy-National Energy Research Scientific Computing Center,
Particle astrophysics research computing center at the University of Maryland,
Institute for Cyber-Enabled Research at Michigan State University,
and Astroparticle physics computational facility at Marquette University;
Belgium {\textendash} Funds for Scientific Research (FRS-FNRS and FWO),
FWO Odysseus and Big Science programmes,
and Belgian Federal Science Policy Office (Belspo);
Germany {\textendash} Bundesministerium f{\"u}r Bildung und Forschung (BMBF),
Deutsche Forschungsgemeinschaft (DFG),
Helmholtz Alliance for Astroparticle Physics (HAP),
Initiative and Networking Fund of the Helmholtz Association,
Deutsches Elektronen Synchrotron (DESY),
and High Performance Computing cluster of the RWTH Aachen;
Sweden {\textendash} Swedish Research Council,
Swedish Polar Research Secretariat,
Swedish National Infrastructure for Computing (SNIC),
and Knut and Alice Wallenberg Foundation;
European Union {\textendash} EGI Advanced Computing for research;
Australia {\textendash} Australian Research Council;
Canada {\textendash} Natural Sciences and Engineering Research Council of Canada,
Calcul Qu{\'e}bec, Compute Ontario, Canada Foundation for Innovation, WestGrid, and Compute Canada;
Denmark {\textendash} Villum Fonden, Carlsberg Foundation, and European Commission;
New Zealand {\textendash} Marsden Fund;
Japan {\textendash} Japan Society for Promotion of Science (JSPS)
and Institute for Global Prominent Research (IGPR) of Chiba University;
Korea {\textendash} National Research Foundation of Korea (NRF);
Switzerland {\textendash} Swiss National Science Foundation (SNSF);
United Kingdom {\textendash} Department of Physics, University of Oxford.

\end{document}